\documentclass[aps,prl,amsmath,two column,amssymb,showpacs]{revtex4}
\usepackage{txfonts}
\usepackage{mathbbold}
\usepackage{epsfig,bm,dcolumn}
\usepackage{graphicx}
\usepackage{color}
\usepackage{overpic}

\begin{document}

\title{BCS-BEC crossover in 2D Fermi gases with Rashba spin-orbit coupling}

\author{Lianyi He}
\email{lianyi@itp.uni-frankfurt.de}

\author{Xu-Guang Huang}
\email{xhuang@itp.uni-frankfurt.de}

\affiliation{ Frankfurt Institute for Advanced Studies and Institute
for Theoretical Physics, J. W. Goethe University, 60438 Frankfurt am
Main, Germany}

\date{\today}

\begin{abstract}
We present a systematic theoretical study of the BCS-BEC crossover
in two-dimensional Fermi gases with Rashba spin-orbit coupling
(SOC). By solving the exact two-body problem in the presence of an
attractive short-range interaction we show that the SOC enhances the
formation of the bound state: the binding energy $E_{\text B}$ and
effective mass $m_{\text B}$ of the bound state grows along with the
increase of the SOC. For the many-body problem, even at weak
attraction, a dilute Fermi gas can evolve from a BCS superfluid
state to a Bose condensation of molecules when the SOC becomes
comparable to the Fermi momentum. The ground-state properties and
the Berezinskii-Kosterlitz-Thouless (BKT) transition temperature are
studied, and analytical results are obtained in various limits. For
large SOC, the BKT transition temperature recovers that for a Bose
gas with an effective mass $m_{\text B}$. We find that the
condensate and superfluid densities have distinct behaviors in the
presence of SOC: the condensate density is generally enhanced by the
SOC due to the increase of the molecule binding, the superfluid
density is suppressed because of the non-trivial molecule effective
mass $m_{\text B}$.
\end{abstract}

\pacs{67.85.Lm, 74.20.Fg, 03.75.Ss, 05.30.Fk}

\maketitle

It has been widely believed for a long time that a smooth crossover
from Bardeen--Cooper--Schrieffer (BCS) superfluidity to
Bose--Einstein condensation (BEC) of molecules could be realized in
an attractive Fermi gas~\cite{Eagles,Leggett,BCSBEC}. This BCS-BEC
crossover phenomenon has been successfully demonstrated in ultracold
fermionic atoms by means of the Feshbach resonance~\cite{BCSBECexp}.
Some recent experimental efforts in generating synthetic non-Abelian
gauge field has opened up the opportunity to study the spin-orbit
coupling (SOC) effect in cold atomic gases~\cite{SOC}. For fermionic
atoms~\cite{SOexp}, it provides an alternative way to study the
BCS-BEC crossover~\cite{3DBCSBEC} according to the theoretical
observation that novel bound states in three dimensions can be
induced by a non-Abelian gauge field even though the attraction is
weak~\cite{3Dbound,NJL}.

Recently, the anisotropic superfluidity in 3D Fermi gases with
Rashba SOC has been intensively studied~\cite{3D,3D2,Zhou}.
Two-dimensional (2D) fermionic systems with Rashba SOC is more
interesting for condensed matter systems~\cite{GR} and topological
quantum computation~\cite{TOP}. By applying a large Zeeman
splitting, a non-Abelian topologically superconducting phase and
Majorana fermionic modes can emerge in spin-orbit coupled 2D
systems~\cite{TOP}. In the absence of SOC, the BCS-BEC crossover and
Berezinskii-Kosterlitz-Thouless (BKT) transition temperature in 2D
attractive fermionic systems were investigated long
ago~\cite{2D,2D2}(see \cite{2Dreview} for a review), which provide a
possible mechanism for pseudogap formation in high-temperature
superconductors~\cite{PG2D}.

In this Letter we present a systematic study of 2D attractive Fermi
gases in the presence of Rashba SOC. The main results are summarized
as follows: (i) The SOC enhances the difermion bound states in 2D.
At large SOC, even for weak intrinsic attraction, the  many-body
ground state is a Bose-Einstein condensate of bound molecules. In
the presence of a harmonic trap, the atom cloud shrinks with
increased SOC. (ii) The BKT transition temperature is enhanced by
the SOC at weak attraction, and for large SOC it tends to the
critical temperature for a gas of molecules with a nontrivial
effective mass. The SOC effect therefore provides a new mechanism
for pseudogap formation in 2D fermionic systems. (iii) In the
presence of SOC, the superfluid ground state exhibits both
spin-singlet and -triplet pairings, and the triplet one has a
non-trivial contribution to the condensate density. In general, the
condensate density is enhanced by the SOC due to the increase of the
molecule binding. However, the superfluid density has entirely
different behavior: it is suppressed by the SOC due to the
increasing molecule effective mass.

\emph{Model and effective potential} --- A quasi-2D Fermi gas can be
realized by arranging a one-dimensional optical lattice along the
axial direction and a weak harmonic trapping potential in the radial
plane, such that fermions are strongly confined along the axial
direction and form a series of pancake-shaped quasi-2D
clouds~\cite{2Dexp,zhang,2Dhu}. The strong anisotropy of the
trapping potentials, namely $\omega_{z}\gg \omega_\bot$ where
$\omega_z$ ($\omega_\bot$) is the axial (radial) frequency, allows
us to use an effective 2D Hamiltonian to deal with the radial
degrees of freedom.

The Hamiltonian of a spin-1/2 attractive Fermi gas with Rashba SOC
is given by $H =\int d^2 {\bf r}\bar{\psi}({\bf r}) \left({\cal
H}_0+\cal{H}_{\rm{so}}\right) \psi({\bf r})-U\int d^2 {\bf
r}\bar{\psi}_{\uparrow}({\bf r})\bar{\psi}_{\downarrow}({\bf
r})\psi_{\downarrow}({\bf r})\psi_{\uparrow}({\bf r})$, where $\psi=
[\psi_\uparrow, \psi_\downarrow]^{\rm T}$ represents the
two-component fermion fields, ${\cal
H}_0=-\frac{\hbar^2\nabla^2}{2m}-\mu-h\sigma_z$ is the free
single-particle Hamiltonian with $\mu$ being the chemical potential
and $h$ the Zeeman splitting, and ${\cal H}_{\rm{so}}
=-i\hbar\lambda(\sigma_x\partial_y-\sigma_y\partial_x)$ is the
Rashba SOC term \cite{sign}. Here $\sigma_{x,y,z}$ are the Pauli
matrices which act on the two-component fermion fields. The short
range attractive interaction is modeled by a contact coupling
$U$~\cite{dilute}. In the following we use the natural units
$\hbar=k_{\text B}=m=1$.

In the functional path integral formalism, the partition function of
the system is ${\cal Z} = \int \mathcal{ D} \psi
\mathcal{D}\bar{\psi}\exp\left\{-{\cal S}[\psi,\bar{\psi}]\right\}$,
where ${\cal S}[\psi,\bar{\psi}]=\int_0^\beta d\tau\left[\int
d^2{\bf r} \bar{\psi}\partial_\tau \psi+H(\psi,\bar{\psi})\right]$
with the inverse temperature $\beta=1/T$. Introducing the auxiliary
complex pairing field
$\Phi(x)=-U\psi_\downarrow(x)\psi_\uparrow(x)$~$[x=(\tau,{\bf r})]$
and applying the Hubbard-Stratonovich transformation, we arrive at
${\cal Z}=\int {\cal D}\Psi{\cal D}\bar{\Psi}{\cal D}\Phi {\cal
D}\Phi^{\ast} \exp\Big\{\frac{1}{2}\int dx\int
dx^\prime\bar{\Psi}(x){\bf
G}^{-1}(x,x^\prime)\Psi(x^\prime)-U^{-1}\int dx|\Phi(x)|^2\Big\}$,
where $\Psi=[\psi,\bar{\psi}]^{\rm T}$ is the Nambu-Gor'kov spinor.
The inverse single-particle Green function ${\bf
G}^{-1}(x,x^\prime)$ is given by
\begin{eqnarray}
{\bf G}^{-1}=\left(\begin{array}{cc}-\partial_{\tau}-{\cal
H}_0-\cal{H}_{\rm{so}} &i\sigma_y\Phi(x)\\  -i\sigma_y\Phi^*(x)&
-\partial_{\tau}+{\cal
H}_0-\cal{H}_{\rm{so}}^{\ast}\end{array}\right)\delta(x-x^\prime).
\end{eqnarray}
Integrating out the fermion fields, we obtain $\mathcal {Z}=\int
\mathcal{D} \Phi \mathcal{D} \Phi^{\ast} \exp \big\{- {\cal
S}_{\rm{eff}}[\Phi, \Phi^{\ast}]\big\}$, where the effective action
reads ${\cal S}_{\rm{eff}}[\Phi, \Phi^{\ast}] = U^{-1}\int dx
|\Phi(x)|^{2} - \frac{1}{2}\mbox{Trln} [{\bf G}^{-1}(x,x^\prime)]$.

\emph{Two-body problem} --- The exact two-body problem at vanishing
density can be studied by considering the Green function $\Gamma(Q)$
of the fermion pairs, where $Q=(i\nu_n,{\bf q})$ with $\nu_n=2n\pi
T$ ($n$ integer) being the bosonic Matsubara frequency. In the
present formalism, $\Gamma^{-1}(Q)$ can be obtained from its
coordinate representation defined as $\Gamma^{-1}(x,x^\prime)=
(\beta V)^{-1}\delta^2{\cal S}_{\rm{eff}}[\Phi, \Phi^{\ast}]/[\delta
\Phi^{\ast}(x)\delta \Phi(x^\prime)]|_{\Phi=0}$. For $\Phi=0$, the
single-particle Green function reduces to its non-interacting form
${\cal G}_0(K)={\rm diag}[g_+(K), g_-(K)]$ with
$g_\pm(K)=[i\omega_n\mp(\xi_{\bf
k}-h\sigma_z)-\lambda(\sigma_xk_y\mp\sigma_yk_x)]^{-1}$, where
$K=(i\omega_n,{\bf k})$ with $\omega_n=(2n+1)\pi T$ being the
fermionic Matsubara frequency. Here $\xi_{\bf k}=\epsilon_{\bf
k}-\mu$ and $\epsilon_{\bf k}={\bf k}^2/2$. The single-particle
spectrum generally has two branches: $\omega_{{\bf k}}^\pm=\xi_{\bf
k}\pm\sqrt{\lambda^2{\bf k}^2+h^2}$.

After the analytical continuation $i\nu_n\rightarrow \omega+i0^+$,
the real part of $\Gamma^{-1}(Q)$ takes the form
\begin{eqnarray}
\Gamma^{-1}(\omega,{\bf q})=\frac{1}{U}-\sum_{\alpha,\gamma=\pm;{\bf
k}}\frac{1-f(\omega_{{\bf k}}^\alpha)-f(\omega_{{\bf
p}}^\gamma)}{4(\omega_{{\bf k}}^\alpha+\omega_{{\bf
p}}^\gamma-\omega)}\left(1+\alpha\gamma {\cal
T}_{\bf{kq}}\right),\label{VEX}
\end{eqnarray}
where $f(E)=1/(e^{\beta E}+1)$ is the Fermi-Dirac distribution
function, and ${\cal T}_{\bf{kq}}=(\lambda^2{\bf k}\cdot{\bf
p}+h^2)/\sqrt{(\lambda^2{\bf k}^2+h^2)(\lambda^2{\bf p}^2+h^2)}$
with ${\bf p}={\bf k}+{\bf q}$. $\Gamma^{-1}$ takes the form similar
to that of the relativistic systems~\cite{He}, due to the fact that
${\cal H}_{\text{so}}$ behaves like a Dirac Hamiltonian. Since in 2D
the bound state forms for arbitrarily small attraction \cite{QM},
the contact coupling $U$ can be regularized by the two-body problem
at vanishing SOC, $U^{-1}=\sum_{\bf k}(2\epsilon_{\bf
k}+\epsilon_{\rm B})^{-1}$ \cite{2D, 2Dreview}, where
$\epsilon_{\text B}$ is the binding energy at vanishing SOC. This
equation recovers the exponential behavior $\epsilon_{\rm
B}=2\Lambda\exp{(-4\pi/U)}$ in 2D~\cite{SW}, where $\Lambda\gg
\epsilon_{\rm B}$ is an energy cutoff. All physical equations are
finally UV convergent in terms of $\epsilon_{\rm B}$ and we set
$\Lambda\rightarrow\infty$ in the dilute limit.

\begin{figure}[!htb]
\begin{center}
\includegraphics[width=9.5cm]{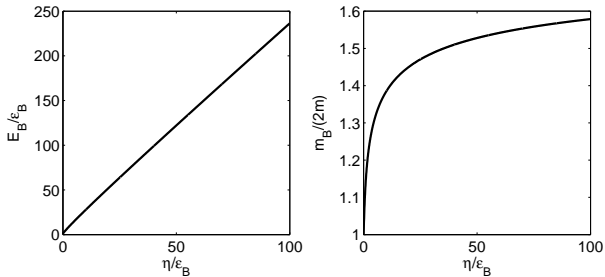}
\caption{The binding energy $E_{\text B}$ (left, divided by
$\epsilon_{\text B}$) and the effective mass $m_{\text B}$ (right,
divided by $2m$) as functions of $\eta/\epsilon_{\text B}$.
 \label{fig1}}
\end{center}
\end{figure}

From now on we consider the case $h=0$. The binding energy $E_{\text
B}$ at nonzero SOC is determined by the solution of
$\omega+2\mu=-E_{\text B}$ for $\Gamma^{-1}(\omega,{\bf q}=0)=0$.
From the imaginary part of $\Gamma^{-1}(Q)$, the bound state
corresponds to the solution in the regime
$-\infty<\omega+2\mu<-\lambda^2$ and hence $E_{\text B}>\lambda^2$.
Completing the momentum integrals analytically, we obtain a simple
algebraic equation for $E_{\text B}$ \cite{supple},
\begin{eqnarray}
\ln\frac{E_{\text B}}{\epsilon_{\text
B}}=\frac{2\lambda}{\sqrt{E_{\text
B}-\lambda^2}}\arctan\frac{\lambda}{\sqrt{E_{\text B}-\lambda^2}}.
\label{EB}
\end{eqnarray}
The solution can be generally expressed as $E_{\text
B}=\epsilon_{\text B}+4\eta J(\eta/\epsilon_{\text B})$ where
$\eta=\lambda^2/2$.  For $\eta\ll\epsilon_{\text B}$, we have
$J\simeq1$ and $E_{\text B}$ is well given by $E_{\text B}\simeq
\epsilon_{\text B}+2\lambda^2$. For $\eta/\epsilon_{\text
B}\rightarrow\infty$, the solution approaches very slowly to the
asymptotic result $E_{\text B}\simeq \lambda^2$. In general,
$E_{\text B}$ increases with increased SOC, as shown in
Fig.\ref{fig1}. It is straightforward to show that the bound state
contains both spin singlet and triplet components~\cite{3Dbound}.

For small nonzero ${\bf q}$, the solution for $\omega$ can be
written as $\omega+2\mu=-E_{\text B}+{\bf q}^2/(2m_{\text B})$,
where $m_{\text B}$ is the molecule effective mass. Substituting
this dispersion into the equation $\Gamma^{-1}(\omega,{\bf q})=0$ we
obtain \cite{supple}
\begin{eqnarray}
\frac{2m}{m_{\text
B}}=1-\frac{1}{2\kappa}\frac{2\sqrt{\kappa-1}-(\kappa-2)(\frac{\pi}{2}-\arctan\frac{\kappa-2}{2\sqrt{\kappa-1}})}{2\sqrt{\kappa-1}
+(\frac{\pi}{2}-\arctan\frac{\kappa-2}{2\sqrt{\kappa-1}})},\label{Mass}
\end{eqnarray}
where $\kappa=E_{\text B}/\lambda^2$. For $\lambda\rightarrow0$, we
obtain the usual result $m_{\text B}\rightarrow2m$. For
$\lambda\rightarrow\infty$, we have $E_{\text
B}\rightarrow\lambda^2$ and $m_{\text B}$ approaches the asymptotic
result $4m$. In general, $m_{\text B}$ is larger than $2m$, as shown
in Fig.\ref{fig1}. Together with the result for $E_{\text B}$, we
conclude that a novel bound state (referred to as rashbon~\cite{3D})
forms. It would have significant impact on the many-body problem
discussed in the following.

\emph{Ground state} --- For the many-body problem, we consider a
homogeneous Fermi gas with fixed fermion density $n=N/V$. For
convenience, we define the Fermi momentum via $n=k_{\text
F}^2/(2\pi)$ and Fermi energy by $\epsilon_{\text F}=k_{\text
F}^2/2$. The ground state ($T=0$) can be studied in the
self-consistent mean-field theory, where we replace the pairing
field $\Phi$ by its expectation value $\langle\Phi\rangle=\Delta$.
Without loss of generality, we set $\Delta$ to be real.

The mean-field ground-state energy $\Omega={\cal S}_{\rm
eff}[\Delta,\Delta]/(\beta V)$ can be evaluated as
$\Omega=\Delta^2/U+(1/2)\sum_{\bf k}(2\xi_{\bf k}-E_{\bf k}^+-E_{\bf
k}^-)$, where $E_{\bf k}^\pm=[(\xi_{\bf k}^\pm)^2+\Delta^2]^{1/2}$
are the quasiparticle excitation energies with $\xi_{\bf
k}^\pm=\xi_{\bf k}\pm\lambda |{\bf k}|$. According to the equation
that $E_{\text B}$ satisfies, $\Omega$ can be evaluated as
$\Omega=\Omega_{\text{2D}}(\Delta,\mu,\epsilon_{\text
B})+\Omega_\lambda$, where
$\Omega_{\text{2D}}(\Delta,\mu,\epsilon_{\text
B})=(\Delta^2/4\pi)\{\ln[(\sqrt{\mu^2+\Delta^2}-\mu)/\epsilon_{\text
B}]-1/2-\mu/(\sqrt{\mu^2+\Delta^2}-\mu)\}$ is formally the
ground-state energy for vanishing SOC~\cite{2D,2Dreview}, and
$\Omega_\lambda=-(\lambda/2\pi)\int_0^\lambda
dk[\sqrt{(\xi_k-\eta)^2+\Delta^2}-(\xi_k-\eta)]$ is the contribution
due to the SOC effect.
\begin{figure}[!htb]
\begin{center}
\includegraphics[width=8.6cm]{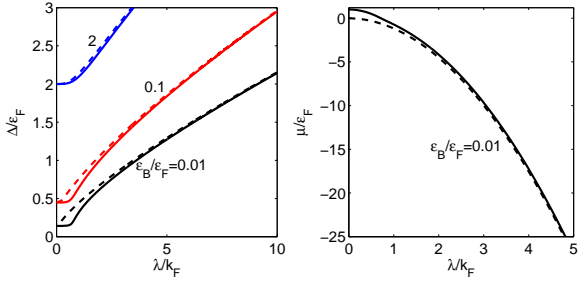}
\caption{(Color-online) The pairing gap $\Delta$ (left, divided by
$\epsilon_{\text F}$) and the chemical potential $\mu$ (right,
divided by $\epsilon_{\text F}$) as functions of $\lambda/k_{\text
F}$.  The dashed lines represents the analytical results
$\Delta=\sqrt{2E_{\text B}\epsilon_{\text F}\zeta(\kappa)}$ and
$\mu=-E_{\text B}/2$ with $E_{\text B}$ calculated from Eq.
(\ref{EB}).
 \label{fig2}}
\end{center}
\end{figure}

From the explicit form of the ground-state energy, the gap and
number equations can be expressed as
\begin{eqnarray}
&&[\mu^2+\Delta^2]^{1/2}-\mu=\epsilon_{\text
B}\exp{\left[2I_1\left(\mu/\eta,\Delta/\eta\right)\right]},\nonumber\\
&&[\mu^2+\Delta^2]^{1/2}+\mu=2\epsilon_{\text
F}-2\eta\left[1-I_2\left(\mu/\eta,\Delta/\eta\right)\right],
\end{eqnarray}
respectively. Here the functions $I_1$ and $I_2$ are defined as
$I_1(a,b)=\int_0^1dx[(x^2-1-a)^2+b^2]^{-1/2}$ and
$I_2(a,b)=\int_0^1dx(x^2-1-a)[(x^2-1-a)^2+b^2]^{-1/2}$. $I_1$, $I_2$
and $\Omega_\lambda$ can be analytically evaluated using the
elliptic functions. For vanishing SOC, we recover the well-known
analytical results, $\Delta=\sqrt{2\epsilon_{\text B}\epsilon_{\text
F}}$ and $\mu=\epsilon_{\text F}-\epsilon_{\text B}/2$~\cite{2D}.

Now let us start from weak attraction, $\epsilon_{\text
B}\ll\epsilon_{\text F}$. For sufficiently small SOC, we have
$I_{\text 1}\rightarrow0$ and $I_{\text 2}\rightarrow-1$, and the
solution is well approximated by $\Delta\simeq\sqrt{2\epsilon_{\text
B}\epsilon_{\text F}}$ and $\mu\simeq\epsilon_{\text
F}-\epsilon_{\text B}/2-2\eta$, which indicates a BCS superfluid
state. For large SOC, we expect that $\mu$ becomes negative and
$|\mu|\gg\Delta$. Substituting this into the gap equation, we find
$\mu\simeq -E_{\text B}/2$, which indicates a Bose-Einstein
condensate of molecules with binding energy $E_{\text B}$. Then
expanding the number equation in powers of $\Delta/|\mu|$ and
keeping the leading order, we obtain $\Delta\simeq\sqrt{2E_{\text
B}\epsilon_{\text F}\zeta(\kappa)}$, where
$\zeta(\kappa)=2\kappa^{-1}(\kappa-1)^{3/2}\big(2\sqrt{\kappa-1}+\frac{\pi}{2}-\arctan\frac{\kappa-2}{2\sqrt{\kappa-1}}\big)^{-1}$.
This is a transparent formula to show that the pairing gap $\Delta$
increases with increased SOC, consistent with the perturbative
approach \cite{Chen}. These analytical results are in good agreement
with the numerical results shown in Fig. \ref{fig2} even for
intermediate $\lambda/k_{\text F}$~\cite{SEB}.

\begin{figure}[!htb]
\begin{center}
\includegraphics[width=9cm]{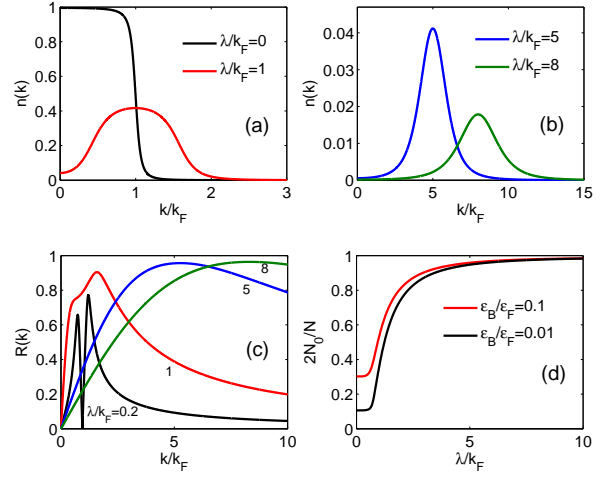}
\caption{(Color-online)(a),(b)\&(c) The momentum distribution $n(k)$
and the ratio
$R(k)=|\phi_{\uparrow\uparrow}(k)|/|\phi_{\uparrow\downarrow}(k)|$
for various values of $\lambda/k_{\text F}$ and $\epsilon_{\text
B}/\epsilon_{\text F}=0.01$. (d) The condensate fraction $2N_0/N$ as
a function of $\lambda/k_{\text F}$ for various values of
$\epsilon_{\text B}/\epsilon_{\text F}$.
 \label{fig3}}
\end{center}
\end{figure}

Using the fermion Green function ${\bf G}(K)$, we can show that the
fermion momentum distribution $n({\bf k})$ is isotropic and can be
expressed as $n(k)=(1/4)\sum_\alpha(1-\xi_k^\alpha/E_k^\alpha)$
\cite{supple}. As shown in Fig. \ref{fig3}, with increased SOC, the
distribution broadens, which indicates a BCS-BEC crossover. The new
feature here is that the distribution generally displays
nonmonotonic behavior. The peak in the distribution is just located
at $k=\lambda$.

The pair wave functions $\phi_{\sigma\sigma^\prime}({\bf
k})\equiv\langle\psi_{{\bf k}\sigma}\psi_{-{\bf
k}\sigma^\prime}\rangle$ can be evaluated as
$\phi_{\uparrow\uparrow}({\bf k})=-(i\Delta/4)e^{i\theta_{\bf
k}}\sum_\alpha\alpha/E_{\bf k}^\alpha$ and
$\phi_{\uparrow\downarrow}({\bf k})=-(\Delta/4) \sum_\alpha1/E_{\bf
k}^\alpha$ \cite{supple}, where $e^{i\theta_{\bf
k}}=(k_x+ik_y)/|{\bf k}|$. Therefore, the superfluid state exhibits
both singlet and triplet pairings for nonzero SOC. The numerical
results for the ratio
$|\phi_{\uparrow\uparrow}(k)|/|\phi_{\uparrow\downarrow}(k)|$
displayed in Fig.\ref{fig3} show that the triplet pairing spreads to
wider momentum regime with increased SOC. According to the general
formula for the condensate number of fermion pairs~\cite{FC},
$N_0=\frac{1}{2}\sum_{\sigma,\sigma^\prime}\int\int d^2{\bf
r}d^2{\bf r}^\prime|\langle\psi_{\sigma}({\bf
r})\psi_{\sigma^\prime}({\bf r}^\prime)\rangle|^2$, the condensate
density reads $n_0=\sum_{\bf k}[|\phi_{\uparrow\downarrow}({\bf
k})|^2+|\phi_{\uparrow\uparrow}({\bf k})|^2]$. The triplet pairing
amplitude contributes, in contrast to the fermionic superfluids with
only singlet pairing~\cite{FCFM}.  For large SOC, we find
analytically that $2N_0/N=1-O(\frac{\Delta^4}{|\mu|^4})\rightarrow
1$ (see also Fig. \ref{fig3}), which indicates the Bose-Einstein
condensation of weakly interacting rashbons.

\begin{figure}[!htb]
\begin{center}
\includegraphics[width=8.6cm]{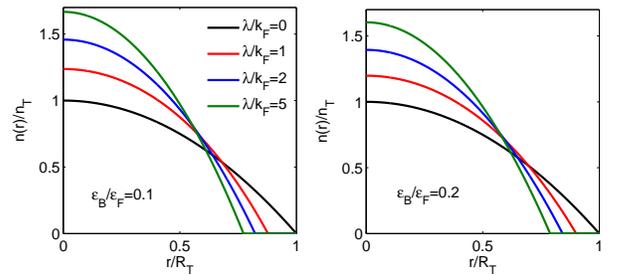}
\caption{(Color-online) The density profile $n(r)$ (divided by
$n_{\text T}=\epsilon_{\text F}/\pi$) in presence of a trap
potential for various values of $\lambda/k_{\text F}$. The Fermi
energy $\epsilon_{\text F}=k_{\text F}^2/2$ in trapped system is
defined as $\epsilon_{\text F}=\sqrt{N}\hbar\omega_\bot$~\cite{He2},
and the Thomas-Fermi radius reads $R_{\text
T}=\sqrt{2\epsilon_{\text F}}/\omega_\bot$.
 \label{fig4}}
\end{center}
\end{figure}

In the presence of a trap potential
$V(r)=\frac{1}{2}\omega_\bot^2r^2$, the chemical potential becomes
$\mu(r)=\mu_0-V(r)$ and the density distribution $n(r)$ can be
solved from the constraint $N=2\pi\int rdr n(r)$ in the local
density approximation. As shown in Fig. \ref{fig4}, the atom cloud
shrinks with increased SOC, which can be viewed as a preliminary
experimental signal of the BCS-BEC crossover.

\begin{figure}[!htb]
\begin{center}
\includegraphics[width=8.7cm]{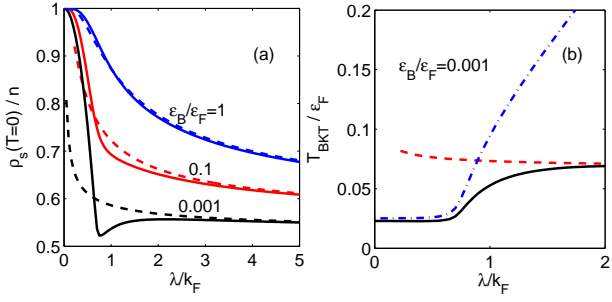}
\caption{(Color-online) (a) The superfluid density $\rho_s$ at $T=0$
(divided by $n$) as a function of $\lambda/k_{\text F}$. The dashed
lines represent the results of $2m/m_{\text B}$ calculated from Eq.
(\ref{Mass}). (b) The BKT transition temperature as a function of
$\lambda/k_{\text F}$. The dashed line represents the rashbon limit
and the dash-dotted line is the mean-field result.
 \label{fig5}}
\end{center}
\end{figure}

\emph{BKT transition temperature} --- At finite temperature in 2D we
should rewrite the complex ordering field $\Phi(x)$ in terms of its
modulus $\Delta(x)$ and phase $\theta(x)$, i.e.,
$\Phi(x)=\Delta(x)\exp[i \theta(x)]$. Since the random fluctuations
of the phase $\theta(x)$ forbid long-range order in 2D, we have
$\langle \Phi(x)\rangle=0$ but $\langle \Delta(x)\rangle\neq0$ at
$T\neq0$. However, Berezinskii~\cite{BKT1} and Kosterlitz and
Thouless~\cite{BKT2} showed that below a critical temperature
$T_{\text{BKT}}$, there exist bound vortex-antivortex pairs and
quasi-long-range order remains.

To determine the BKT transition temperature, we derive an effective
action for the U$(1)$ phase field $\theta(x)$. To this end we make a
gauge transformation $\psi(x)=\exp{[i\theta(x)/2]}
\chi(x)$~\cite{2D2,2Dreview}. Then we arrive at the expression $
{\cal Z}=\int\Delta\mathcal{D}\Delta\mathcal{D}\theta
\exp{\big\{-\beta{\cal
U}_{\text{eff}}[\Delta(x),\partial\theta(x)]\big\}}$, where the
effective action $\beta{\cal U}_{\text{eff}}[\Delta(x),
\partial\theta(x)] = U^{-1}\int dx\Delta^2(x) -
\frac{1}{2}\mbox{Trln} {\bf S}^{-1}[\Delta(x),\partial\theta(x)]$
now depends on the modulus-phase variables. The Green function of
the initial (charged) fermions takes a new form ${\bf
S}^{-1}[\Delta(x),\partial\theta(x)]={\cal
G}^{-1}[\Delta(x)]-\Sigma[\partial \theta(x)]$. Here ${\cal
G}^{-1}[\Delta(x)]={\bf G}^{-1}[\Delta(x),\Delta(x)]$ is the green
function of the neutral fermion, and $\Sigma[\partial \theta] \equiv
\tau_{3}[i\partial_{\tau}\theta/2+(\nabla\theta)^{2}/8]-\hat{I}[i
\nabla^{2}\theta/4+i\nabla\theta\cdot\nabla/2]+(\lambda/2)[\tau_3\sigma_x\partial_y\theta
-\hat{I}\sigma_y\partial_x\theta]$, where $\tau_i (i=1,2,3)$ are the
Pauli matrices in the Nambu-Gor'kov space.

Since the low-energy dynamics for $\Delta\neq 0$ is governed by
long-wavelength fluctuations of $\theta(x)$, we neglect the
amplitude fluctuations and treat $\Delta$ as its saddle point
value~\cite{2D2,2Dreview}. Then the effective action can be
decomposed as ${\cal U}_{\text{eff}}[\Delta(x),
\partial \theta(x)] \simeq {\cal U}_{\rm kin} [\Delta,
\partial \theta(x)] + {\cal U}_{\rm pot} (\Delta)$. The potential
part reads ${\cal U}_{\rm pot}/V=\Delta^2/U+\sum_{\bf k}[\xi_{\bf
k}-{\cal W}(E_{\bf k}^+)-{\cal W}(E_{\bf k}^-)]$ where ${\cal
W}(E)=E/2+T\ln(1+e^{-\beta E})$. The kinetic part can be obtained by
the derivative expansion $\beta{\cal U}_{\rm kin} [\Delta,
\partial \theta(x)] =
\sum_{n=1}^{\infty}\frac{1}{n} \mbox{Tr}(\mathcal{G} \Sigma)^{n}$.

Keeping only lowest-order derivatives of $\theta(x)$, we find that
the kinetic term ${\cal U}_{\text{kin}}$ coincides with the
classical spin XY-model, which has the continuum Hamiltonian
$H_{\mathrm{XY}} = \frac{1}{2}{\cal J}\int d^2{\bf r}\, [\nabla
\theta ({\bf r})]^{2}$ where the phase stiffness ${\cal
J}=\frac{\rho_s}{4m}$ and $\rho_s$ is the superfluid
density~\cite{Naoto}. The superfluid density in our model can be
evaluated as $\rho_s=n-\rho_1-\rho_2$, where
$\rho_1=(\lambda/8\pi)\sum_{\alpha=\pm}\int_0^\infty dk
\alpha(\xi_k^\alpha+\Delta^2/\xi_k)[1-2f(E_k^\alpha)]/E_k^\alpha$
and $\rho_2=-(1/4\pi)\sum_{\alpha=\pm}\int_0^\infty
kdk(k+\alpha\lambda)^2f^\prime(E_k^\alpha)$ \cite{supple}. The BKT
transition temperature is determined by
$T_{\text{BKT}}=\frac{\pi}{2}{\cal J}$~\cite{BKT1,BKT2,Naoto,note}.

For sufficiently small $\epsilon_{\text B}$ and SOC, $\Delta$ is
correspondingly small and $T_{\text{BKT}}$ recovers the mean-field
result $T_\Delta$. On the other hand, for large $\epsilon_{\text B}$
and/or SOC, $\rho_s$ can be well approximated by its
zero-temperature value for $T\sim T_{\text{BKT}}$. We are interested
in the case with small $\epsilon_{\text B}$ and large SOC. For large
SOC, using the fact $\Delta\ll|\mu|$, we find analytically that
\cite{supple}
\begin{eqnarray}
\rho_s(T\ll T_\Delta)\simeq \frac{2m}{m_{\text B}}n,\ \ \ \ {\cal
J}(T\ll T_\Delta)\simeq\frac{n_{\text B}}{m_{\text B}},
\end{eqnarray}
where $n_{\text B}=n/2$ and $m_{\text B}$ is given by Eq.
(\ref{Mass}). Therefore, the phase stiffness ${\cal J}$ naturally
recovers that for a  Bose (rashbon) gas at large SOC. The BKT
transition temperature and the phase stiffness jump $\Delta {\cal
J}$ reaches the rashbon limit $T_{\text{BKT}}=\pi n_{\text
B}/(2m_{\text B})=(2m/m_{\text B})\epsilon_{\text F}/8$ and $\Delta
{\cal J}=n_{\text B}/m_{\text B}$. To verify above analytical
results, we show the numerical results for $\rho_s(T=0)$ and
$T_{\text{BKT}}$ in Fig. \ref{fig5}. Even for weak attraction, a
visible pseudogap phase appears in the window
$T_{\text{BKT}}<T<T_{\Delta}$ for $\lambda\sim k_{\rm F}$. The SOC
therefore provides a new mechanism for pseudogap formation in 2D
fermionic systems.

Finally, we point out a surprising  result, $\rho_s<n$ at $T=0$,
which is in contrast to the result $\rho_s=n$ for fermionic
superfluids in the absence of SOC~\cite{Naoto,RHOS}. Actually, at
$T=0$, the superfluid density reads $\rho_s=n-\rho_\lambda$, where
the $\lambda$-dependent term $\rho_\lambda=\rho_1(T=0)$ is always
positive and is generally an increasing function of $\lambda$.
Therefore, the superfluid density shown in Fig. \ref{fig3} has
entirely different behavior in contrast to the condensate density
shown in Fig. \ref{fig5}: It is generally suppressed by the SOC
effect. The exact two-body solution provides a very transparent
explanation to this suppression. At large SOC, the effective mass
$m_{\text B}>2m$ is an increasing function of SOC and causes the
suppression of the superfluid density by a factor $2m/m_{\text B}$.
Our argument also applies to the suppression of the radial ($x-y$
plane) superfluid density $\rho_s^\perp$ for the 3D
case~\cite{Zhou}, where the radial effective mass $m_{\text
B}^\perp$ is larger than $2m$~\cite{3D}.

\emph{Acknowledgments} --- L. He acknowledges the support from the
Alexander von Humboldt Foundation, and X.-G. Huang is supported by
the Deutsche Forschungsgemeinschaft (Grant SE 1836/1-2).

{\bf Note Added} --- After finishing this Letter, we note that
similar results of the condensate density \cite{Zhou, Wan} and the
superfluid density \cite{Zhou} in spin-orbit coupled Fermi gases are
also reported.

\newpage

\begin{widetext}

{\it Appendix:} In this supplementary material, we present the
derivation details of some results in the main text.

\subsection {(A) Two-Body Problem: Binding Energy and Effective Mass}
Using the free fermion propagators $g_\pm(K)$, $\Gamma^{-1}(Q)$ can
be expressed as
\begin{eqnarray}
\Gamma^{-1}(Q)=\frac{1}{U}-\frac{1}{2}\sum_K\text{Tr}\left[g_+(K+Q)\sigma_yg_-(K)\sigma_y\right].\label{A1}
\end{eqnarray}
Completing the Matsubara frequency sum, we obtain Eq. (2) of the
text. For the two-body problem, we discard the Fermi-Dirac
distribution function and define the solution for
$\Gamma^{-1}(\omega,{\bf q})=0$ as $E_{\bf q}=-(\omega+2\mu)$. The
two-body equation becomes
\begin{eqnarray}
\sum_{\bf k}\left(\frac{2}{k^2+\epsilon_{\rm B}}-\frac{2{\cal
E}_{\bf kq}}{{\cal E}_{\bf
kq}^2-4\lambda^2k^2-\frac{4\lambda^4k^2q^2\sin^2\varphi}{{\cal
E}_{\bf kq}^2-\lambda^2q^2}}\right)=0.\label{A2}
\end{eqnarray}
Here $\varphi$ is the angle between ${\bf k}$ and ${\bf q}$, and
${\cal E}_{\bf kq}=E_{\bf q}+\epsilon_{{\bf k}+{\bf
q}/2}+\epsilon_{{\bf k}-{\bf q}/2}=E_{\bf q}+k^2+q^2/4$.

For zero center-of-mass momentum ${\bf q}$, the above equation
reduces to $\int_0^\infty kdk[2(k^2+\epsilon_{\rm
B})^{-1}-\sum_{\alpha=\pm}(k^2+2\alpha\lambda k+E_{\text
B})^{-1}]=0$. The integral can be carried out directly. The easiest
way is to use the trick $k^2\pm 2\lambda k=(k\pm
\lambda)^2-\lambda^2$. Since the integrals are logarithmically
divergent, we can convert the integration variables to
$k\pm\lambda$. Finally we find that it becomes
\begin{eqnarray}
\int_0^\infty dz\left(\frac{1}{z+\epsilon_{\rm B}}-
\frac{1}{z+E_{\text
B}}\right)-2\lambda\int_0^\lambda\frac{dk}{k^2+E_{\text
B}-\lambda^2}=0.\label{A3}
\end{eqnarray}
Using the condition $E_{\text B}>\lambda^2$ we then obtain Eq. (3)
of the text.

For nonzero center-of-mass momentum ${\bf q}$, we write $E_{\text
q}\simeq E_{\text B}-{\bf q}^2/(2m_{\text B})$ for small $q^2$ and
expand Eq. (\ref{A2}) to the order $O(q^2)$, then we obtain
\begin{eqnarray}
\left(1-\frac{2m}{m_{\text B}}\right)\int_0^\infty
kdk\frac{(k^2+E_{\text B})^2+4\lambda^2k^2}{\left[(k^2+E_{\text
B})^2-4\lambda^2k^2\right]^2}=\int_0^\infty
kdk\frac{8\lambda^4k^2}{(k^2+E_{\text B})\left[(k^2+E_{\text
B})^2-4\lambda^2k^2\right]^2}.\label{A4}
\end{eqnarray}
Defining $\kappa=E_{\text B}/\lambda^2$, this equation becomes
\begin{eqnarray}
1-\frac{2m}{m_{\text B}}=\int_0^\infty
dx\frac{8x}{(x+\kappa)[(x+\kappa)^2-4x]^2}\left[\int_0^\infty dx
\frac{(x+\kappa)^2+4x}{[(x+\kappa)^2-4x]^2}\right]^{-1}.\label{A5}
\end{eqnarray}
Completing the integrals analytically, we obtain Eq. (4) of the
text.

\subsection {(B) Derivation of the Ground-State Energy}
In the mean-field approximation, the ground-state energy can be
expressed as
\begin{eqnarray}
\Omega=\frac{\Delta^2}{U}-\frac{1}{2}\frac{1}{\beta}\sum_n\sum_{\bf
k}\text{lndet}{\cal G}^{-1}(i\omega_n,{\bf k}),\label{A6}
\end{eqnarray}
where the inverse fermion Green function reads
\begin{eqnarray}
{\cal G}^{-1}(i\omega_n,{\bf
k})=\left(\begin{array}{cc}i\omega_n-\xi_{\bf k}+h\sigma_z-\lambda(k_y\sigma_x-k_x\sigma_y)&i\sigma_y\Delta\\
-i\sigma_y\Delta& i\omega_n+\xi_{\bf
k}-h\sigma_z-\lambda(k_y\sigma_x+k_x\sigma_y)\end{array}\right).\label{A7}
\end{eqnarray}
Using the formula for block matrix, we first work out the
determinant and obtain
\begin{eqnarray}
\text{det}{\cal G}^{-1}(i\omega_n,{\bf
k})=\left[(i\omega_n)^2+h^2-\xi_{\bf
k}^2-\lambda^2k^2-\Delta^2\right]^2-4h^2(i\omega_n)^2-4\lambda^2k^2\left(\xi_{\bf
k}^2-h^2\right).\label{A8}
\end{eqnarray}
Then completing the Matsubara frequency sum and taking $T=0$ we
obtain $\Omega=\Delta^2/U+(1/2)\sum_{\bf k}(2\xi_{\bf k}-E_{\bf
k}^+-E_{\bf k}^-)$ where the term $\sum_{\bf k}\xi_{\bf k}$ is added
to recover the correct ground state energy for the normal state
($\Delta=0$). The quasiparticle dispersions are given by the
positive roots of the equation $\det{\cal G}^{-1}=0$, i.e.,
\begin{eqnarray}
E_{\bf k}^\pm=\left[\xi_{\bf
k}^2+\Delta^2+\lambda^2k^2+h^2+2\sqrt{\xi_{\bf
k}^2(\lambda^2k^2+h^2)+h^2\Delta^2}\right]^{1/2}.\label{A9}
\end{eqnarray}
For $h=0$, they reduces to $E_{\bf k}^\pm=\sqrt{(\xi_{\bf
k}\pm\lambda k)^2+\Delta^2}$. At finite temperature, the
thermodynamic potential reads $\Omega={\cal U}_{\rm
pot}/V=\Delta^2/U+\sum_{\bf k}[\xi_{\bf k}-{\cal W}(E_{\bf
k}^+)-{\cal W}(E_{\bf k}^-)]$ where ${\cal
W}(E)=E/2+T\ln(1+e^{-\beta E})$.

For $T=0$ and $h=0$, the ground-state energy can be expressed in
terms of $E_{\text B}$ as
$\Omega=(\Delta^2/4\pi)\sum_{\alpha=\pm}\int_0^\infty
kdk[(2\epsilon_k+2\alpha \lambda k+E_{\text
B})^{-1}-(E_k^\alpha+\xi_k^\alpha)^{-1}]$. Since the integrals are
convergent, we can use the trick $k^2\pm 2\lambda k=(k\pm
\lambda)^2-\lambda^2$ and convert the integration variables to
$k\pm\lambda$. After a straightforward calculation, we obtain
\begin{eqnarray}
\Omega=\Omega_{\text{2D}}(\Delta,\mu,E_{\text
B})+\frac{\Delta^2}{4\pi}\frac{2\lambda}{\sqrt{E_{\text
B}-\lambda^2}}\arctan\frac{\lambda}{\sqrt{E_{\text
B}-\lambda^2}}+\Omega_\lambda.\label{A10}
\end{eqnarray}
Noticing the fact that $E_{\text B}$ satisfies Eq. (3) of the text,
we obtain $\Omega=\Omega_{\text{2D}}(\Delta,\mu,\epsilon_{\text
B})+\Omega_\lambda$.

\subsection {(C) Solution of the Gap and Number Equations at Large SOC}
The original forms of the gap and number equations at $T=0$ are
\begin{eqnarray}
&&\frac{1}{U}=\frac{1}{2}\sum_{\bf k}\left(\frac{1}{2E_{\bf
k}^+}+\frac{1}{2E_{\bf k}^-}\right),\ \ \ \ \ n=\sum_{\bf
k}\left(1-\frac{\xi_{\bf k}^+}{2E_{\bf k}^+}-\frac{\xi_{\bf
k}^-}{2E_{\bf k}^-}\right).\label{A11}
\end{eqnarray}
For large SOC, we expect $\mu<0$ and $\Delta\ll|\mu|$. Therefore, we
can expand the equations in powers of $\Delta/|\mu|$ and keep only
the leading order terms. The gap equation becomes
\begin{eqnarray}
\int_0^\infty kdk\left(\frac{2}{k^2+\epsilon_{\rm
B}}-\sum_{\alpha=\pm}\frac{1}{k^2+2\alpha\lambda k-2\mu}\right)=0.
\label{A12}
\end{eqnarray}
We obtain $\mu=-E_{\text B}/2$. Substituting this into the number
equation, we obtain
\begin{eqnarray}
n=\frac{\epsilon_{\text
F}}{\pi}=\frac{\Delta^2}{2\pi}\sum_{\alpha=\pm}\int_0^\infty
kdk\frac{1}{(k^2+2\alpha\lambda k+E_{\text
B})^2}=\frac{\Delta^2}{\pi}\int_0^\infty kdk \frac{(k^2+E_{\text
B})^2+4\lambda^2k^2}{\left[(k^2+E_{\text
B})^2-4\lambda^2k^2\right]^2}. \label{A12}
\end{eqnarray}
We notice that the integral also appears in Eq. (\ref{A4}).
Completing the integral analytically, we obtain
$\Delta=\sqrt{2E_{\text B}\epsilon_{\text F}\zeta(\kappa)}$ where
$\zeta(\kappa)$ is defined in the text.

\subsection {(D) The Fermion Green Function and Related Quantities}
The explicit form of the fermion Green function ${\cal
G}(i\omega_n,{\bf k})$ can be evaluated using the formula for block
matrix. For $h=0$, we find that the matrix elements (in the
Nambu-Gor'kov space) can be expressed as
\begin{eqnarray}
&&{\cal G}_{11}={\cal A}_{11}+\frac{k_y\sigma_x-k_x\sigma_y}{k}{\cal
B}_{11},\ \ \ \ {\cal
G}_{22}={\cal A}_{22}+\frac{k_y\sigma_x+k_x\sigma_y}{k}{\cal B}_{22},\nonumber\\
&&{\cal G}_{12}=-i\sigma_y\left[{\cal
A}_{12}+\frac{k_y\sigma_x+k_x\sigma_y}{k}{\cal B}_{12}\right],\ \ \
\ {\cal G}_{21}=i\sigma_y\left[{\cal
A}_{21}+\frac{k_y\sigma_x-k_x\sigma_y}{k}{\cal
B}_{21}\right].\label{A13}
\end{eqnarray}
Here ${\cal A}_{ij}$ and ${\cal B}_{ij}$ take the forms
\begin{eqnarray}
&&{\cal A}_{11}=\frac{1}{2}\sum_{\alpha=\pm}\frac{i\omega_n+\xi_{\bf
k}^\alpha}{(i\omega_n)^2-(E_{\bf k}^\alpha)^2},\ \ \ \ {\cal
A}_{22}=\frac{1}{2}\sum_{\alpha=\pm}\frac{i\omega_n-\xi_{\bf
k}^\alpha}{(i\omega_n)^2-(E_{\bf k}^\alpha)^2},\nonumber\\
&&{\cal
A}_{12}=\frac{1}{2}\sum_{\alpha=\pm}\frac{\Delta}{(i\omega_n)^2-(E_{\bf
k}^\alpha)^2},\ \ \ \ {\cal A}_{21}={\cal A}_{12},\label{A14}
\end{eqnarray}
and
\begin{eqnarray}
&&{\cal
B}_{11}=\frac{1}{2}\sum_{\alpha=\pm}\alpha\frac{i\omega_n+\xi_{\bf
k}^\alpha}{(i\omega_n)^2-(E_{\bf k}^\alpha)^2},\ \ \ \ {\cal
B}_{22}=-\frac{1}{2}\sum_{\alpha=\pm}\alpha\frac{i\omega_n-\xi_{\bf
k}^\alpha}{(i\omega_n)^2-(E_{\bf k}^\alpha)^2},\nonumber\\
&&{\cal
B}_{12}=-\frac{1}{2}\sum_{\alpha=\pm}\alpha\frac{\Delta}{(i\omega_n)^2-(E_{\bf
k}^\alpha)^2},\ \ \ \ {\cal B}_{21}=-{\cal B}_{12}.\label{A15}
\end{eqnarray}

Using the matrix elements of the Green function, we can calculate
various quantities. First, the momentum distribution can be
evaluated as
\begin{eqnarray}
n({\bf k})\equiv\langle\bar{\psi}_{{\bf k}\uparrow}\psi_{{\bf
k}\uparrow}\rangle=\langle\bar{\psi}_{{\bf k}\downarrow}\psi_{{\bf
k}\downarrow}\rangle=\frac{1}{\beta}\sum_n{\cal
A}_{11}(i\omega_n,{\bf k})e^{i\omega_n0^+}.\label{A16}
\end{eqnarray}
Second, the singlet and triplet pairing amplitudes can be expressed
as
\begin{eqnarray}
&&\phi_{\uparrow\downarrow}({\bf k})\equiv\langle\psi_{{\bf
k}\uparrow}\psi_{-{\bf
k}\downarrow}\rangle=\frac{1}{\beta}\sum_n{\cal
A}_{21}(i\omega_n,{\bf k}),\ \ \ \ \phi_{\downarrow\uparrow}({\bf
k})\equiv\langle\psi_{{\bf k}\downarrow}\psi_{-{\bf
k}\uparrow}\rangle=-\frac{1}{\beta}\sum_n{\cal
A}_{21}(i\omega_n,{\bf
k}),\nonumber\\
&&\phi_{\uparrow\uparrow}({\bf k})\equiv\langle\psi_{{\bf
k}\uparrow}\psi_{-{\bf
k}\uparrow}\rangle=-\frac{k_y-ik_x}{k}\frac{1}{\beta}\sum_n{\cal
B}_{21}(i\omega_n,{\bf k}),\ \ \ \ \phi_{\downarrow\downarrow}({\bf
k})\equiv\langle\psi_{{\bf k}\downarrow}\psi_{-{\bf
k}\downarrow}\rangle=\frac{k_y+ik_x}{k}\frac{1}{\beta}\sum_n{\cal
B}_{21}(i\omega_n,{\bf k}).\label{A17}
\end{eqnarray}
Therefore, we have the relations $\phi_{\uparrow\downarrow}({\bf
k})=-\phi_{\downarrow\uparrow}({\bf k})$ and
$\phi_{\uparrow\uparrow}({\bf
k})=-\phi_{\downarrow\downarrow}^*({\bf k})$.

According to Leggett's definition~\cite{FC}, the condensate number
of fermion pairs is given by
\begin{eqnarray}
N_0=\frac{1}{2}\sum_{\sigma,\sigma^\prime=\uparrow,\downarrow}\int\int
d^2{\bf r}d^2{\bf r}^\prime|\langle\psi_{\sigma}({\bf
r})\psi_{\sigma^\prime}({\bf r}^\prime)\rangle|^2. \label{A18}
\end{eqnarray}
For systems with only singlet pairing, this recovers the usual
result $N_0=\int\int d^2{\bf r}d^2{\bf
r}^\prime|\langle\psi_{\uparrow}({\bf r})\psi_{\downarrow}({\bf
r}^\prime)\rangle|^2$. Converting this to the momentum space, we
find that the condensate density $n_0=N_0/V$ should be a sum of all
absolute squares of the pairing amplitudes. The final result for
$T=0$ is
\begin{eqnarray}
n_0&=&\frac{1}{2}\sum_{\bf k}\left[|\phi_{\uparrow\downarrow}({\bf
k})|^2+|\phi_{\downarrow\uparrow}({\bf
k})|^2+|\phi_{\uparrow\uparrow}({\bf
k})|^2+|\phi_{\downarrow\downarrow}({\bf k})|^2\right]\nonumber\\
&=&\frac{1}{8}\sum_{\bf k}\left[\frac{\Delta^2}{(E_{\bf
k}^+)^2}+\frac{\Delta^2}{(E_{\bf k}^-)^2}\right].\label{A19}
\end{eqnarray}
For large attraction and/or SOC, we expect $\Delta\ll|\mu|$. Using
the number equation (\ref{A11}) and expanding all terms in powers of
$\Delta/|\mu|$, we can show that $2N_0/N=1-O(\Delta^4/|\mu|^4)$.
Therefore, the condensate fraction approaches unity at large
attraction and/or SOC.

\subsection {(E) Effective Action of the Phase Field}
To obtain the effective action for the phase field $\theta(x)$ to
the order $(\nabla\theta)^2$, we notice that the available operators
in $\Sigma[\partial\theta]$ are $\Sigma_1=\tau_3(\nabla\theta)^2/8$,
$\Sigma_2=-\hat{I}\nabla\theta\cdot\nabla/2$ and
$\Sigma_3=(\lambda/2)[\tau_3\sigma_x\partial_y\theta-\hat{I}\sigma_y\partial_x\theta]$.
According to the derivative expansion, we have carefully checked
that there are four types of nonzero contributions:
\begin{eqnarray}
{\cal U}_1\sim\text{Tr}({\cal G}\Sigma_1),\ \ \ {\cal U}_2\sim
\text{Tr}({\cal G}\Sigma_2{\cal G}\Sigma_2),\ \ \ {\cal U}_3\sim
\text{Tr}({\cal G}\Sigma_3{\cal G}\Sigma_3),\ \ \ {\cal U}_4\sim
\text{Tr}({\cal G}\Sigma_2{\cal G}\Sigma_3).\label{A21}
\end{eqnarray}
Since the superfluid state is isotropic, the phase stiffness should
also be isotropic. We have carefully checked that all anisotropic
terms vanish exactly. Completing the trace in the Nambu-Gor'kov and
spin spaces, we finally obtain the following expressions for the
four types of contributions:
\begin{eqnarray}
&&{\cal U}_1=\frac{1}{2}\left[\frac{1}{\beta}\sum_n\sum_{\bf
k}\frac{1}{4}\left({\cal A}_{11}e^{i\omega0^+}-{\cal
A}_{22}e^{-i\omega_n0^+}\right)\right]\int d^2{\bf
r}(\nabla\theta)^2\nonumber\\
&&{\cal U}_2=\frac{1}{2}\left[\frac{1}{\beta}\sum_n\sum_{\bf
k}\frac{k^2}{8}\left({\cal A}_{11}^2+{\cal B}_{11}^2+{\cal
A}_{22}^2+{\cal B}_{22}^2+2{\cal A}_{21}^2+2{\cal
B}_{21}^2\right)\right]\int d^2{\bf
r}(\nabla\theta)^2,\nonumber\\
&&{\cal U}_3=\frac{1}{2}\left[\frac{1}{\beta}\sum_n\sum_{\bf
k}\frac{\lambda^2}{4}\left({\cal A}_{11}^2+{\cal A}_{22}^2+2{\cal
A}_{21}^2\right)\right]\int d^2{\bf
r}(\nabla\theta)^2,\nonumber\\
&&{\cal U}_4=\frac{1}{2}\left[\frac{1}{\beta}\sum_n\sum_{\bf
k}\frac{\lambda k}{2}\left({\cal A}_{11}{\cal B}_{11}-{\cal
A}_{22}{\cal B}_{22}+2{\cal A}_{21}{\cal B}_{21}\right)\right]\int
d^2{\bf r}(\nabla\theta)^2.\label{A22}
\end{eqnarray}
Collecting all terms, the effective action is reduced to a spin
XY-model Hamiltonian $H_{\text{XY}}=\frac{1}{2}{\cal J}\int d^2{\bf
r}[\nabla\theta({\bf r})]^2$, where the phase stiffness ${\cal J}$
is given by
\begin{eqnarray}
&&{\cal J}=\frac{1}{\beta}\sum_n\sum_{\bf
k}\Bigg[\frac{1}{4}\left({\cal A}_{11}e^{i\omega0^+}-{\cal
A}_{22}e^{-i\omega_n0^+}\right)+\frac{k^2}{8}\left({\cal
A}_{11}^2+{\cal B}_{11}^2+{\cal A}_{22}^2+{\cal B}_{22}^2+2{\cal
A}_{21}^2+2{\cal
B}_{21}^2\right)\nonumber\\
&&\ \ \ \ \ +\frac{\lambda^2}{4}\left({\cal A}_{11}^2+{\cal
A}_{22}^2+2{\cal A}_{21}^2\right)+\frac{\lambda k}{2}\left({\cal
A}_{11}{\cal B}_{11}-{\cal A}_{22}{\cal B}_{22}+2{\cal A}_{21}{\cal
B}_{21}\right)\Bigg].\label{A23}
\end{eqnarray}
Completing the Matsubara frequency sum we then obtain the expression
given in the text.

\subsection {(F) Properties of the Superfluid Density}
First, setting $\Delta=0$, we find that $\rho_s=0$. Therefore
$\rho_s$ vanishes exactly in the normal state, as expected. Second,
for vanishing SOC, the expressions of $\rho_s$ and ${\cal J}$
recover the well known form given in \cite{2Dreview}. Here we will
examine the behavior of $\rho_s$ for large SOC at $T=0$. At zero
temperature, the superfluid density reduces to
\begin{eqnarray}
\rho_s=n-\rho_\lambda,\ \ \
\rho_\lambda=\frac{\lambda}{8\pi}\int_0^\infty
dk\left[\left(\xi_k^++\frac{\Delta^2}{\xi_k}\right)\frac{1}{E_k^+}-\left(\xi_k^-+\frac{\Delta^2}{\xi_k}\right)\frac{1}{E_k^-}\right].\label{A24}
\end{eqnarray}
Therefore, even at $T=0$, the superfluid stiffness does not recover
the result $\rho_s=n$ for ordinary fermionic superfluids. Let us
show what happens at large $\lambda$. In this case $\mu\simeq
-E_{\text{B}}/2$ and $\Delta\ll|\mu|$. Therefore, we can expand the
expression in powers of $\Delta/|\mu|$ and keep only the leading
order terms. Doing so, we obtain (see Eq. (\ref{A12}))
\begin{eqnarray}
n\simeq\frac{\Delta^2}{8\pi}\lambda\int_0^\infty
kdk\left[\frac{1}{(\xi_k^+)^2}+\frac{1}{(\xi_k^-)^2}\right]
\simeq\frac{\Delta^2}{\pi}\int_0^\infty kdk \frac{(k^2+E_{\text
B})^2+4\lambda^2k^2}{\left[(k^2+E_{\text
B})^2-4\lambda^2k^2\right]^2},\label{A25}
\end{eqnarray}
and
\begin{eqnarray}
\rho_\lambda&\simeq&\frac{\Delta^2}{8\pi}\lambda\int_0^\infty
dk\left\{\frac{1}{\xi_k}\left(\frac{1}{\xi_k^+}-\frac{1}{\xi_k^-}\right)-\frac{1}{2}\left[\frac{1}{(\xi_k^+)^2}-\frac{1}{(\xi_k^-)^2}\right]\right\}\nonumber\\
&\simeq&\frac{\Delta^2}{\pi}\int_0^\infty kdk
\frac{8\lambda^4k^2}{(k^2+E_{\text B})\left[(k^2+E_{\text
B})^2-4\lambda^2k^2\right]^2}.\label{A26}
\end{eqnarray}
Comparing the above results with Eq. (\ref{A4}), we find that
$\rho_\lambda/n=1-2m/m_{\text B}$. Therefore, for large SOC, the
superfluid density and the phase stiffness are reduced to
\begin{eqnarray}
\rho_s=\frac{2m}{m_{\text B}}n,\ \ \ \ {\cal J}=\frac{2m}{m_{\text
B}}\frac{n}{4m}=\frac{n_{\text B}}{m_{\text B}}\label{A27}
\end{eqnarray}
where $n_{\text B}=n/2$ is the density of rashbons. This means that,
at large SOC, the phase stiffness self-consistently recovers that
for a rashbon gas.

\end{widetext}


\begin{thebibliography}{99}
\bibitem{Eagles}      {D. M. Eagles, Phys. Rev. {\bf 186}, 456(1969).}
\bibitem{Leggett}     {A. J. Leggett, in \emph{Modern trends in the theory of condensed matter}, Springer-Verlag, Berlin, 1980, pp.13-27.}
\bibitem{BCSBEC}      {P. Nozieres and S. Schmitt-Rink, J. Low Temp. Phys. {\bf 59}, 195(1985);
                       C. A. R. S¡äa de Melo \emph{et. al.}, Phys. Rev. Lett. {\bf 71}, 3202(1993).}
\bibitem{BCSBECexp}   {M. Greiner \emph{et al.}, Nature {\bf 426}, 537(2003);
                       S. Jochim \emph{et al.}, Science {\bf 302}, 2101(2003);
                       M. W. Zwierlein \emph{et al.}, Nature {\bf 435}, 1047(2003).}
\bibitem{SOC}         {K. Osterloh \emph{et al.}, Phys. Rev. Lett. {\bf 95}, 010403(2005);
                       J. Ruseckas \emph{et al.}, Phys. Rev. Lett. {\bf 95}, 010404(2005);
                       T. D. Stanescu \emph{et al.}, Phys. Rev. Lett. {\bf 99}, 110403 (2007);
                       X. J. Liu \emph{et al.}, Phys. Rev. Lett. {\bf 102}, 046402(2009);
                       Y. J. Lin \emph{et al.}, Nature {\bf 462}, 628(2009);
                       Y. J. Lin \emph{et al.}, Nature {\bf 471}, 83(2011).}
\bibitem{SOexp}       {J. D. Sau \emph{et al.}, Phys. Rev. {\bf B83}, 140510(R) (2011).}
\bibitem{3DBCSBEC}    {J. P. Vyasanakere \emph{et al.}, Phys. Rev. {\bf B84}, 014512 (2011).}
\bibitem{3Dbound}     {J. P. Vyasanakere and V. B. Shenoy, Phys. Rev. {\bf B83}, 094515 (2011).}
\bibitem{NJL}         {This phenomenon is analogous to the catalysis of the dynamical mass generation by an external
                       non-Ablelian field in quantum field theory, see V. P. Gusynin \emph{et al.}, Phys. Rev. {\bf D57}, 5230 (1998);
                       I. A. Shovkovy and V. M. Turkowski, Phys. Lett. \bf{B367}, 213 (1996).}
\bibitem{3D}          {H. Hu \emph{et al.}, Phys. Rev. Lett. {\bf 107}, 195304(2011);
                       Z. -Q. Yu and H. Zhai, Phys. Rev. Lett. {\bf 107}, 195305(2011);}
\bibitem{3D2}         {M. Iskin and A. L. Subasi, Phys. Rev. Lett. {\bf 107}, 050402(2011);
                       M. Gong, \emph{et al.}, Phys. Rev. Lett. {\bf 107}, 195303(2011);
                       W. Yi and G. -C. Guo, Phys. Rev. {\bf A84}, 031608(R) (2011);
                       L. Han and C. A. R. S¡äa de Melo, Phys. Rev. {\bf A85}, 011606(R) (2012);
                       L. Dell'Anna \emph{et al.}, Phys. Rev. {\bf A84}, 033633(2011).}
\bibitem{Zhou}        {K. Zhou and Z. Zhang, Phys. Rev. Lett. {\bf 108}, 025301 (2012).}
\bibitem{GR}          {L. P. Gor'kov and E. I. Rashba, Phys. Rev. Lett. {\bf 87}, 037004 (2001).}
\bibitem{TOP}         {C. Zhang \emph{et al.}, Phys. Rev. Lett. {\bf 101}, 160401 (2008);
                       J. D. Sau \emph{et al.}, Phys. Rev. {\bf B82}, 214509 (2010);
                       S. Tewari \emph{et al.}, New J. Phys. {\bf 13}, 065004 (2011).}
\bibitem{2D}          {M. Randeria \emph{et al.}, Phys. Rev. Lett. {\bf 62}, 981 (1989);
                       Phys. Rev. {\bf B41}, 327(1990). }
\bibitem{2D2}         {V. P. Gusynin \emph{et al.}, JETP {\bf 88}, 685(1999);
                       JETP {\bf 90}, 993(2000).}
\bibitem{2Dreview}    {V. M. Loktev \emph{et al.}, Phys. Rept. {\bf 349}, 1 (2001).}
\bibitem{PG2D}        {The experimental observation of pairing pseudogap in two-dimensional Fermi gases has been recently
                       reported in M. Feld \emph{et al.}, Nature {\bf 480}, 75(2011).}
\bibitem{2Dexp}       {S. Stock \emph{et al.}, Phys. Rev. Lett. {\bf 95}, 190403 (2005);
                       Z. Hadzibabic \emph{et al.}, Nature {\bf 441}, 1118 (2006).}
\bibitem{zhang}       {W. Zhang \emph{et al.},  Phys. Rev. {\bf A77}, 063613 (2008).}
\bibitem{2Dhu}        {P. Dyke \emph{et al.}, Phys. Rev. Lett. {\bf 106}, 105304 (2011). }
\bibitem{sign}        {The sign of the coupling constant $\lambda$ is not important, since all physical quantities
                       depends only on $\lambda^2$. In this paper we set $\lambda>0$ without loss of generality.}
\bibitem{dilute}      {The validity of a contact interaction is restricted in the dilute limit, i.e.,
                       $k_{\text F}r_0\ll1$, where $k_{\text F}$ is the Fermi momentum defined through
                       the fermion density $n=k_{\text F}^2/(2\pi)$ and $r_0$ is the effective range of the
                       attractive interaction. In presence of Rashba SOC, another dilute condition $\lambda r_0\ll 1$
                       should also be fulfilled, see X. Cui, Phys. Rev. {\bf A85}, 022705 (2012).}
\bibitem{He}          {L. He and P. Zhuang, Phys. Rev. {\bf D75}, 096003 (2007); Phys. Rev. {\bf D76}, 056003 (2007);
                       G. Sun, \emph{et al.}, Phys. Rev. {\bf D75}, 096004 (2007).}
\bibitem{QM}          {L. D. Landau and E. M. Lifshitz, \emph{Quantum Mechanics---Non Relativistic Theory} (Pergamon Press, New York, 1989).}
\bibitem{SW}          {For an inter-atomic potential described by a 2D circularly symmetric well of radius
                       $r_0$ and depth $\upsilon_0$, the binding energy $\epsilon_{\rm B}$ is given by
                       $\epsilon_{\rm B} = 1/(2r_0^2) \exp[-2/(\upsilon_0r^2_0)]$ in the dilute limit $\upsilon_0r^2_0\rightarrow0$ \cite{QM}.
                       For quasi-2D cold atoms confined by an axial trapping frequency $\omega_z$, the binding energy is given by
                       $\epsilon_{\text B}=(C\hbar\omega_z/\pi)\exp[\sqrt{2\pi}l_z/a_s]$, where $a_s$ is the 3D s-wave scattering length,
                       $l_z = \sqrt{\hbar/\omega_z}$, and $C\simeq0.915$. See D. S. Petrov and G. V. Shlyapnikov, Phys. Rev. {\bf A64}, 012706 (2001).}
\bibitem{supple}      {See Supplemental Material for details of the derivation.}
\bibitem{Chen}        {G. Chen \emph{et al.}, Phys. Rev. {\bf A85}, 013601 (2012).}
\bibitem{SEB}         {Actually, for large $\epsilon_{\text B}/\epsilon_{\text F}$, the analytical formulas work well even for
                       small SOC since $\Delta\ll|\mu|$ can be easily satisfied.}
\bibitem{FC}          {A. J. Leggett, \emph{Quantum Liquids. Bose Condensation and Cooper Pairing in Condensed-Matter Systems} (Oxford Universty Press, Oxford, 2006).}
\bibitem{FCFM}        {L. Salasnich \emph{et al.}, Phys. Rev. {\bf A72}, 023621(2005);
                       L. Salasnich, Phys. Rev. {\bf A76}, 015601(2007).}
\bibitem{He2}         {L. He and P. Zhuang, Phys. Rev. {\bf A78}, 033613 (2008).}
\bibitem{BKT1}        {V. L. Berezinskii, Sov. Phys. JETP {\bf 32}, 493 (1971).}
\bibitem{BKT2}        {J. M. Kosterlitz and D. Thouless, J. Phys. {\bf C5}, L124 (1972).}
\bibitem{Naoto}       {N. Nagaosa, \emph{Quantum Field Theory in Condensed Matter Physics}, (Springer, 1999).}
\bibitem{note}        {Since ${\cal J}$ depends on the variables $\Delta,\mu$ and $T$ explicitly, this equation should be
                       accompanied with the gap equation $\partial{\cal U}_{\text{pot}}/\partial\Delta=0$ and the number equation
                       $-\partial{\cal U}_{\text{pot}}/\partial\mu=N$~\cite{2D2,2Dreview}.}
\bibitem{RHOS}        {E. Taylor \emph{et al.}, Phys. Rev. {\bf A74}, 063626(2006);
                       N. Fukushima, \emph{et al.}, Phys. Rev. {\bf A75}, 033609(2007).}
\bibitem{Wan}         {B. Huang and S. Wan,  arXiv:1109.3970.}
\end{thebibliography}
\end{document}